\documentstyle[preprint,aps,epsfig]{revtex}
\tightenlines
\newcommand{\bea}{\begin{eqnarray}}
\newcommand{\eea}{\end{eqnarray}}
\newcommand{\beq}{\begin{equation}}
\newcommand{\eeq}{\end{equation}}
\newcommand{\bay}{\begin{array}}
\newcommand{\eay}{\end{array}}
\newcommand{\vslash}{\mbox{$\not{\hspace{-1.03mm}v}$}}        

\newcommand{\epsslash}{\mbox{$\not{\hspace{-0.5mm}\varepsilon}$}}
\begin{document}
\preprint{\parbox{6cm}{\flushright UCSD/PTH 00-03\\[1cm]}}
\title{Long-distance effects in $B\to V\gamma$ 
radiative weak decays}
\author{Benjamin Grinstein and Dan Pirjol\\[1cm]}
\address{ Department of Physics,
University of California at San Diego, La Jolla, CA 92093}
\date{\today}
\maketitle

\begin{abstract}
A systematic approach to long-distance effects in exclusive radiative
weak $B$ decays is presented, based on a combination of the heavy quark
limit with perturbative QCD. The dominant long-distance effects,
connected with weak annihilation and $W$-exchange topologies, can
be computed in a model-independent way using experimental data on
$B$ radiative leptonic decays. Nonfactorizable corrections vanish
in the chiral limit and to leading twist. 
The left-handed photon amplitudes are shown to be enhanced relative to 
the right-handed ones, both in the long- and short-distance parts of 
the $\bar B$ decay amplitudes.
Recent CLEO data on $B\to K^{*}\gamma$ decays are consistent with 
Standard Model estimates of the long-distance contributions, and
disfavor an enhanced gluonic penguin contribution.
We discuss the implications of our results for the 
extraction of $|V_{td}|$. 
\end{abstract}

\pacs{pacs1,pacs2,pacs3}

\narrowtext

\section{Introduction}

Rare radiative decays of $B$ mesons received considerable theoretical
attention due to their special sensitivity to physics beyond the
Standard Model. Alternatively, if the validity of the SM is taken
for granted, these decays can offer useful information on the magnitude
of the CKM parameters. So far, most of the theoretical effort has 
been concentrated on aspects of inclusive decays $b\to s\gamma$, 
which can be treated with the help of an operator product expansion 
(OPE) combined with the heavy quark expansion \cite{Greub}. 
The corresponding strong interaction effects can be quantified in 
terms of a few nonperturbative matrix elements.
Experimental results on the branching ratio for inclusive 
$B\to X_s\gamma$ decays \cite{CLEO1,CLEO2} appear to be in agreement 
with the predictions of the Standard Model.

On the other hand, the treatment of exclusive radiative decays such 
as $B\to K^*\gamma$ or $B\to \rho\gamma$ is considerably more 
difficult. This is due to the fact that bound state effects must
be taken into account, which are essentially nonperturbative in nature.
In addition to the matrix element of the penguin operator
(short-distance component), a nonlocal contribution must be included
too, arising from a product of the electromagnetic
coupling with the weak nonleptonic Hamiltonian (long-distance
component). These effects have been estimated using various methods, 
such as perturbative QCD combined with the quark model
\cite{GSW,pQCD,ABS,EIM,NLL}, dispersion methods \cite{DoGoPe},
vector meson dominance \cite{VMD,HYC} and QCD sum rules
\cite{LD1,LD1.5,LD2}. 

Recently the CLEO collaboration measured the branching ratios for
several exclusive radiative decays $B\to V\gamma$, with 
$V = K^{*0}, K^{*+}$ \cite{CLEOexcl}, and gave upper limits on the 
$V=\rho,\omega$ modes. A more detailed study of these decays is
therefore well motivated.

In this paper we propose a different approach to the calculation of 
the dominant long distance contributions to $B\to V\gamma$ decays
(with $V$ a member of the SU(3) octet of vector mesons), 
arising from weak annihilation and $W$-exchange topologies. We start
by parametrizing all these decays in terms of a few common amplitudes
(for each photon helicity $\lambda=L,R$), using SU(3)
symmetry. It can be shown that the structure of these universal 
amplitudes simplifies very much to the leading order of a twist 
expansion, in powers of $\Lambda/E_\gamma$ and $\Lambda/p_+ = 
\Lambda/m_B$, with $p_+$ the light-cone component of the vector meson 
momentum. Since the photon energy in $B\to\rho(K^*)\gamma$ 
decays is $E_\gamma = 2.6$ GeV, the higher twist corrections
can be expected to be well under control. We find that, to leading
twist, the photons emitted in $\bar B$ decay are predominantly 
left-handed.
Second, the nonfactorizable corrections turn out to be computable in 
terms of the light-cone wavefunctions of the $B$ and $V$ mesons. 
Furthermore, in the chiral limit, the nonfactorizable corrections
vanish exactly to one-loop order. The remaining factorizable 
contribution can be determined in a model-independent way 
using experimental data on radiative leptonic $B$ decays $B\to
\gamma e\nu$.

These results can be used to assess the feasibility of the
determination of the ratio $|V_{td}/V_{ts}|$ from a comparison
of the $B\to K^*\gamma$ and $B\to \rho\gamma$ decays. 
The Appendix describes certain constraints on radiative
decay form-factors following from a Ward identity expressing the 
conservation of the electromagnetic current.

\section{Flavor SU(3) predictions for $\bar B\to V\gamma$ decay 
amplitudes}

In the Standard Model, the radiative $\bar B\to V\gamma$ decays 
(with $V$ a member of the SU(3) octet of vector mesons) are 
dominated by the operators ($q=s,d$) (see, e.g., \cite{BuBuLa})
\bea\label{Hpeng}
{\cal H}_{rad} &=& \frac{-4G_F}{\sqrt2} \lambda_t^{(s)} 
\sum_{i=7,8}C_i(\mu) {\cal O}^{(s)}_i(\mu)  + (s\to d)\,,\\
{\cal O}^{(q)}_7 &=& \frac{e}{16\pi^2} F^{\mu\nu}
\bar q\sigma_{\mu\nu}(m_b P_R + m_q P_L) b\,,\\
{\cal O}^{(q)}_8 &=& \frac{g_s}{16\pi^2}
\bar q\sigma_{\mu\nu}(m_b P_R + m_q P_L) G^{\mu\nu}b\,,\quad
P_{R,L}=\frac12(1\pm\gamma_5)
\eea
We denote here and in the following the combination of CKM matrix 
elements $\lambda_q^{(q')} = V_{qb} V_{qq'}^*$. In addition to this,
there are also contributions mediated by the usual weak nonleptonic
Hamiltonian 
\bea\label{Hweak}
{\cal H}_W &=& \frac{4G_F}{\sqrt2} \left\{ \lambda_u^{(s)}
[ C_1 (\bar u b)_{V-A}(\bar s u)_{V-A}  +
C_2 (\bar s b)_{V-A}(\bar u u)_{V-A}]\right.\\
&+&\left. \lambda_c^{(s)}
[ C_1 (\bar c b)_{V-A}(\bar s c)_{V-A}  +
C_2 (\bar s b)_{V-A}(\bar c c)_{V-A}] -
\lambda_t^{(s)}\sum_{i=3}^{10}c_i Q_i^{(s)} + (s\to d)
\right\}\nonumber
\eea
with $(\bar q_1 q_2)_{V-A} (\bar q_3 q_4)_{V-A} = 
[\bar q_1\gamma_\mu P_L q_2][\bar q_3\gamma^\mu P_L q_4]$. 
In these diagrams, the photon attaches to the
internal quark lines with the usual electromagnetic coupling. We 
will neglect in the following the contributions of the penguin 
operators $Q_{3-10}$, which are suppressed relative to those of the 
current-current operators by their smaller Wilson coefficients.

In a quark diagram language, all $B\to V\gamma_\lambda$ decay 
amplitudes are parametrized by nine independent $SU(3)$ amplitudes, 
for each individual photon helicity $\lambda = L,R$. They
appear together in fewer combinations, such that a certain 
predictive power is still retained. They include penguin-type 
amplitudes
$P_{u,c,t}$ -- corresponding to topologies with a $u$-, $c$- and 
$t$-quark running in the loop, the weak annihilation $(WA)$-type 
amplitude $A$, the $W$-exchange amplitude $E$, penguin-annihilation 
amplitudes $PA$,
and amplitudes with one insertion of the gluonic penguin operator
${\cal O}_8$ which will be denoted
$M_i$ (see Fig.~1). In the penguin $P^{(i)}_{u,c}$ and $M^{(i)}$
amplitudes we distinguish between diagrams with the photon coupling
to the loop quark or the other emerging light quark $(i=1)$, and
diagrams with the photon coupling to the spectator quark $(i=2)$.
We adopt the usual phase conventions for the vector states $(\rho^+,
\rho^0,\rho^-) = (u\bar d,\frac{1}{\sqrt2}(d\bar d-u\bar u),
-d\bar u)$, $(\bar K^{*0}, K^{*-})=(s\bar d,-s\bar u)$,
$(K^{*+}, K^{*0})=(u\bar s,d\bar s)$, $\phi^{(8)}=\frac{1}{\sqrt6}
(2s\bar s-u\bar u-d\bar d)$, and for the heavy mesons
$(B^+, B^0, B_s) = (\bar bu, \bar bd, \bar bs)$,
$(B^-, \bar B^0, \bar B_s) = (b\bar u, -b\bar d, b\bar s)$.

The weak annihilation amplitudes $A^{(i)}$ contribute only to 
the $B^\pm$ radiative decays, where they appear always in the same
combination $A\equiv \frac23 A^{(1)} + \frac23 A^{(2)} -
\frac13 A^{(3)}$ (see Fig.~1(a)). Writing explicitly the 
contributions of the penguin amplitudes one has
\bea
& &A(B^-\to \rho^-\gamma_\lambda) =
\lambda_u^{(d)} (P^{(1)}_{u\lambda} + Q_u P^{(2)}_{u\lambda} 
+ A_\lambda) +
\lambda_c^{(d)} (P^{(1)}_{c\lambda} + Q_u P^{(2)}_{c\lambda})\\
& &\qquad\qquad
+ \lambda_t^{(d)} (P_{t\lambda} + M^{(1)}_\lambda + Q_u M^{(2)}_\lambda)
\nonumber\\
& &A(B^-\to K^{*-}\gamma_\lambda) = \lambda_u^{(s)} 
(P^{(1)}_{u\lambda} + Q_u P^{(2)}_{u\lambda} 
+ A_\lambda) + 
\lambda_c^{(s)} (P^{(1)}_{c\lambda} + Q_u P^{(2)}_{c\lambda})\label{6}\\
& &\qquad\qquad
 + \lambda_t^{(s)} (P_{t\lambda}+ M^{(1)}_\lambda + Q_u M^{(2)}_\lambda)
\,.\nonumber
\eea

The penguin amplitude with an internal $t$-quark 
$P_{t\lambda}$
(arising from the penguin operator ${\cal O}_7$) is usually
called the
short-distance contribution, in contrast to the long-distance 
amplitude,
arising from nonlocal insertions of the weak Hamiltonian 
(\ref{Hweak}) with one photon attachment.
In a hadronic language, the penguin amplitudes $P^{(i)}_u$ receive
contributions from rescattering effects of the form $B^\pm\to\{\rho^\pm
\rho^0\}\to \rho^\pm\gamma$, and the charmed penguins $P^{(i)}_c$ arise
 from rescattering processes $B^\pm\to\{D^{*\pm} \bar D^{*0}\},
\{\psi\rho^0\}\to \rho^\pm\gamma$.

The $\bar B^0$ decay amplitudes contain the $W$-exchange topology,
parametrized by the graphical amplitudes $E^{(i)}$ (Fig.~1(b)).
All the decays discussed below contain these amplitudes in the
same combination $E\equiv \frac23 E^{(2)} + \frac23 E^{(3)} -
\frac13 E^{(1)}$
\bea\label{7}
& &A(\bar B^0\to \rho^0\gamma_\lambda) =
\frac{1}{\sqrt2}
\lambda_u^{(d)} (P^{(1)}_{u\lambda} + Q_d P^{(2)}_{u\lambda} 
 - E_\lambda - (Q_u-Q_d)PA_{u\lambda})\\
& &\qquad +
\frac{1}{\sqrt2} \lambda_c^{(d)} (P^{(1)}_{c\lambda} + 
Q_d P^{(2)}_{c\lambda} - (Q_u-Q_d)PA_{c\lambda})
 + \frac{1}{\sqrt2} \lambda_t^{(d)} 
(P_{t\lambda} + M^{(1)}_\lambda + Q_d M^{(2)}_\lambda)\nonumber\\
& &A(\bar B^0\to \bar K^{*0}\gamma_\lambda) = \lambda_u^{(s)} 
(P^{(1)}_{u\lambda} + Q_d P^{(2)}_{u\lambda})
 + \lambda_c^{(s)} (P^{(1)}_{c\lambda} + Q_d P^{(2)}_{c\lambda})
+ \lambda_t^{(s)} (P_{t\lambda}+ M^{(1)}_\lambda + Q_d M^{(2)}_\lambda)\label{8}\\
& &A(\bar B^0\to \phi^{(8)}\gamma_\lambda) =
-\frac{1}{\sqrt6}\lambda_u^{(d)}(P^{(1)}_{u\lambda}
+ Q_d P^{(2)}_{u\lambda} + E_\lambda + (Q_u+Q_d-2Q_s)PA_{u\lambda}) \\
& &\qquad
-\frac{1}{\sqrt6}\lambda_c^{(d)}(P^{(1)}_{c\lambda} + Q_d
P^{(2)}_{c\lambda}+(Q_u+Q_d-2Q_s)PA_{c\lambda})
-\frac{1}{\sqrt6}\lambda_t^{(d)}
(P_{t\lambda}+ M^{(1)}_\lambda + Q_d M^{(2)}_\lambda)\,. \nonumber
\eea

Finally, the corresponding $\bar B_s$ decays are written in terms of the
amplitudes previously introduced as
\bea
& &A(\bar B_s\to K^{*0}\gamma_\lambda) =
-\lambda_u^{(d)} (P^{(1)}_{u\lambda} + Q_s P^{(2)}_{u\lambda} )
-\lambda_c^{(d)} (P^{(1)}_{c\lambda} + Q_s P^{(2)}_{c\lambda} )\\
& &\qquad\qquad\qquad\qquad -
\lambda_t^{(d)} (P_{t\lambda}+ M^{(1)}_\lambda + Q_s M^{(2)}_\lambda)\nonumber\\
& &A(\bar B_s\to \rho^0\gamma_\lambda) = \frac{1}{\sqrt2}
\lambda_u^{(s)} (E_\lambda+(Q_u-Q_d)PA_{u\lambda}) + \frac{1}{\sqrt2}
\lambda_c^{(s)}(Q_u-Q_d)PA_{c\lambda}\\
& &A(\bar B_s\to \phi^{(8)}\gamma_\lambda) = -\frac{1}{\sqrt6}
\lambda_u^{(s)}
\left( 2P^{(1)}_{u\lambda} +2Q_s P^{(2)}_{u\lambda}
-  E_\lambda - (Q_u+Q_d-2Q_s)PA_{u\lambda}\right)\label{12}\\
& &\qquad
-\frac{1}{\sqrt6} \lambda_c^{(s)}(2P^{(1)}_{c\lambda} +
2Q_s P^{(2)}_{c\lambda} - (Q_u+Q_d-2Q_s)PA_{c\lambda})
- \sqrt{\frac23}\lambda_t^{(s)}
(P_{t\lambda}+ M^{(1)}_\lambda + Q_s M^{(2)}_\lambda)\,.\nonumber
\eea
Note that the long-distance penguin-type amplitudes with an
internal $u$ and $c$-quark are different in $B^\pm$ and $B^0,
B_s$ decays, due to the different electric charge of the spectator
quark in the two cases. This is different from the conclusion 
reached in \cite{ABS}; present experimental data from CLEO (see 
Eqs.~(\ref{exp1}), (\ref{exp2}) below) appear to confirm the presence
of these long-distance contributions at the $1\sigma$ level.

Decays to the SU(3) singlet vector meson
$\phi^{(1)}=-\frac{1}{\sqrt3}(u\bar u+d\bar d+s\bar s)$ introduce 
new amplitudes $S_\lambda$, which arise from diagrams similar to $PA_\lambda$
(see Fig.~1(d)) but with the photon attaching to the quark in the loop or 
to the spectator in the $\bar B$ meson
\bea
A(\bar B^0\to \phi^{(1)}\gamma_\lambda) &=& 
-\frac{1}{\sqrt3}\lambda_u^{(d)}
(P_{u\lambda}^{(1)}+Q_d P_{u\lambda}^{(2)} + E_\lambda +S_{u\lambda})\\
& &\qquad
-\frac{1}{\sqrt3}\lambda_c^{(d)}(P_{c\lambda}^{(1)}+Q_d P_{c\lambda}^{(2)}
+S_{c\lambda}) 
-\frac{1}{\sqrt3}\lambda_t^{(d)}(P_{t\lambda}^{(1)} +
M_{\lambda}^{(1)}+Q_d M_{\lambda}^{(2)})\nonumber\\
A(\bar B_s\to \phi^{(1)}\gamma_\lambda) &=& 
\frac{1}{\sqrt3}\lambda_u^{(s)}
(P_{u\lambda}^{(1)}+Q_d P_{u\lambda}^{(2)} + E_\lambda +S_{u\lambda})\\
& &\qquad
+\frac{1}{\sqrt3}\lambda_c^{(s)}(P_{c\lambda}^{(1)}+Q_d P_{c\lambda}^{(2)}
+S_{c\lambda}) 
+\frac{1}{\sqrt3}\lambda_t^{(s)}(P_{t\lambda}^{(1)} +
M_{\lambda}^{(1)}+Q_d M_{\lambda}^{(2)})\nonumber\,.
\eea

The decay amplitudes into the physical states $\phi$, $\omega$
are obtained by combining the octet and singlet amplitudes with a
mixing angle $\theta_V$ as
\bea
\phi &=& \phi^{(8)}\cos\theta_V - \phi^{(1)}\sin\theta_V\\
\omega &=& \phi^{(8)}\sin\theta_V + \phi^{(1)}\cos\theta_V\,.
\eea
The physical value of the mixing angle is close to $\tan\theta_V = 
\frac{1}{\sqrt2}$, corresponding to ideal mixing $\phi=\bar ss$,
$\omega=-\frac{1}{\sqrt2}(u\bar u+d\bar d)$. Combining the amplitudes for decays
into SU(3) eigenstates one finds the following amplitudes for decay into
physical states $\omega, \phi$
\bea
A(\bar B^0\to \omega\gamma_\lambda) &=& 
-\frac{1}{\sqrt2}\lambda_u^{(d)}
(P_{u\lambda}^{(1)}+Q_d P_{u\lambda}^{(2)} + E_\lambda +\frac13 PA_{u\lambda}
+ \frac23 S_{u\lambda})\\
& &
-\frac{1}{\sqrt2}\lambda_c^{(d)}(P_{c\lambda}^{(1)}+Q_d P_{c\lambda}^{(2)}
+\frac13 PA_{c\lambda} + \frac23 S_{c\lambda}) 
-\frac{1}{\sqrt2}\lambda_t^{(d)}(P_{t\lambda}^{(1)} +
M_{\lambda}^{(1)}+Q_d M_{\lambda}^{(2)})\nonumber\\
A(\bar B^0\to \phi\gamma_\lambda) &=&
\frac13\lambda_u^{(d)}(-PA_{u\lambda} + S_{u\lambda}) +
\frac13\lambda_c^{(d)}(-PA_{c\lambda} + S_{c\lambda})\,,
\eea
and
\bea
A(\bar B_s\to \omega\gamma_\lambda) &=& 
\frac{1}{\sqrt2}\lambda_u^{(s)}
( E_\lambda +\frac13 PA_{u\lambda}
+ \frac23 S_{u\lambda})
+\frac{1}{\sqrt2}\lambda_c^{(s)}
(\frac13 PA_{c\lambda} + \frac23 S_{c\lambda}) \\
A(\bar B_s\to \phi\gamma_\lambda) &=&
-\lambda_u^{(s)}(P_{u\lambda}^{(1)} + Q_s P_{u\lambda}^{(2)}
-\frac13 PA_{u\lambda} + \frac13 S_{u\lambda})\\
& & -
\lambda_c^{(s)}(
P_{c\lambda}^{(1)} + Q_s P_{c\lambda}^{(2)}
-\frac13 PA_{c\lambda} + \frac13 S_{c\lambda})
-\lambda_t^{(s)}(
P_{t\lambda} + M^{(1)}_\lambda + Q_s M^{(2)}_\lambda)
\,.\nonumber
\eea
The radiative decay widths are given, in terms of these amplitudes,
by
\beq
\Gamma (\bar B\to V\gamma) = \frac{E_\gamma}{8\pi m_B^2}
\sum_{\lambda=L,R} |A(\bar B\to V\gamma_\lambda)|^2\,.
\eeq

\subsection{The short-distance amplitude}

Due to the smallness of the light quark masses $m_d$, $m_s$
appearing in the short-distance Hamiltonian (\ref{Hpeng}), relative to
the $b$-quark mass, it couples predominantly to left-handed
photons. The amplitude for emitting right-hand polarized photons
is suppressed relative to the one for left-handed photons 
by the quark mass ratio $|P_{tR}/P_{tL}| = {\cal O}(m_{d,s}/m_b)$.
Therefore, the right-handed amplitude $P_{tR}$ can be neglected to a 
good approximation. 

The left-handed short-distance amplitude can be expressed 
in terms of the $B\to V$ form factors of the tensor current, defined 
as
\bea
\langle V(p',\epsilon)|\bar q\sigma_{\mu\nu} b|\bar B(p)\rangle &=&
g_+(q^2) \varepsilon_{\mu\nu\lambda\sigma} \epsilon^*_\lambda
(p+p')_\sigma + 
g_-(q^2) \varepsilon_{\mu\nu\lambda\sigma} 
\epsilon^*_\lambda (p-p')_\sigma\\
 & &+
h(q^2) \varepsilon_{\mu\nu\lambda\sigma} (p+p')_\lambda
(p-p')_\sigma(\epsilon^*\cdot p)\nonumber\,.
\eea
Using the relation $\sigma_{\mu\nu}\gamma_5 = \frac{i}{2}
\varepsilon_{\mu\nu\alpha\beta}\sigma^{\alpha\beta}$ (corresponding
to $\varepsilon^{0123}=1$), one finds
\bea\nonumber
& &\langle V(p',\epsilon)|\bar q\sigma_{\mu\nu}\gamma_5 b|
\bar B(p)\rangle  =\\
& &\qquad
-ig_+(q^2)[\epsilon^*_\mu (p+p')_\nu - \epsilon^*_\nu (p+p')_\mu] 
-ig_-(q^2)[\epsilon^*_\mu (p-p')_\nu - \epsilon^*_\nu (p-p')_\mu]
\nonumber\\
& &\qquad -ih(q^2)(\epsilon^*\cdot p)
[(p+p')_\mu (p-p')_\nu - (p+p')_\nu (p-p')_\mu]\,.
\eea
The relevant quantity is the formfactor $g_+$ at the kinematical
point $q^2=0$. This has been computed using light-cone QCD sum rules
(LCSR) \cite{LCSR,BB} and lattice QCD \cite{UKQCD}, with results 
which are in reasonable agreement with each other. The most recent 
LCSR calculation gave $g_+^{(\rho)}(0)=0.29\pm 0.04$
and $g_+^{(K^*)}(0)=0.38\pm 0.06$ \cite{BB}, which compares well
with the lattice result $g_+^{(K^*)}(0)=0.32^{+0.04}_{-0.02}$
\cite{UKQCD}. A somewhat larger value
$g_+^{(K^*)}(0)=0.4$ was recently extracted \cite{LiWi} from the 
$D\to K^*e\nu$ semileptonic form factors using heavy
quark symmetry relations \cite{IsWi}.

Using the latter value for $g_+(0)$, one finds for the short-distance
penguin amplitudes with an internal $t$ quark
\bea
P_{tL} &=& -8\sqrt{2} G_F \frac{e}{16\pi^2}
C_7(m_b) m_b m_B E_\gamma g_+(q^2=0)\\
 &=& (1.8\times 10^{-6})
\left(\frac{g_+(0)}{0.4}\right)
\left(\frac{m_b\mbox{ (GeV)}}{4.2}\right)\mbox{ GeV}\nonumber
\eea
and $P_{tR}\simeq 0$.
We used in this numerical estimate the leading-log value for the 
Wilson coefficient $C_7(m_b)=-0.31$. 
In the following section we will prove a similar suppression of the 
right-handed amplitude relative to the left-handed one for the weak 
annihilation ($A$) and the $W$-exchange ($E$) amplitudes in $\bar B$
decays, to leading 
order in an expansion in powers of $1/E_\gamma$.

\subsection{Weak annihilation and $W$-exchange amplitudes}

We start by assuming factorization; the corrections to this 
approximation will be discussed below.
The factorized weak annihilation amplitude contributing to the
$B^-\to\rho^-\gamma$ decay is written as
\bea\label{11}
A_\lambda = \frac{G_F}{\sqrt2} a_1 \left\{
-f_B p_{B\mu} 
\langle \rho^-\gamma_\lambda |(\bar du)_{V-A}^\mu|0\rangle
+ m_{\rho^-} f_{\rho^-} (\epsilon^*_\rho)_\mu\langle \gamma_\lambda |
(\bar ub)_{V-A}^\mu|B^-\rangle \right\}\,,
\eea
corresponding to the photon coupling to the $B^-$ or to the $\rho^-$
constituent quarks respectively. To tree level, the Wilson
coefficient $a_1$ is given by $a_1 = C_1(m_b) + C_2(m_b)/N_c = 1.02$
\cite{BuBuLa}. The first term can be computed exactly in the chiral
limit with the result
\bea\label{first}
p_{B\mu} \langle \rho^-\gamma_\lambda |(\bar du)_{V-A}^\mu|0\rangle =
-\langle\rho^-\gamma_\lambda |i\partial_\mu (\bar du)_{V-A}^\mu|0\rangle
= -em_{\rho^-} f_{\rho^-}(\epsilon^*_\rho\cdot\epsilon^*_\lambda) +
{\cal O}(m_u,m_d)\,.
\eea
We used here the following relations for the divergence of the
isovector and isoaxial currents in the presence of an 
electromagnetic field ${\cal A}$ (with $e > 0$)
\bea
i\partial_\mu (\bar q\gamma_\mu \lambda^i q) &=& -e {\cal A}_\lambda
\bar q [\hat {\cal Q}, \lambda^i]\gamma_\lambda q -
\bar q [\hat m, \lambda^i] q\\
i\partial_\mu (\bar q\gamma_\mu\gamma_5 \lambda^i q) &=& 
-e {\cal A}_\lambda
\bar q [\hat {\cal Q}, \lambda^i]\gamma_\lambda\gamma_5 q -
\bar q \{\hat m, \lambda^i\}\gamma_5 q\,.
\eea
$\hat {\cal Q}=\mbox{diag }(\frac23,-\frac13,-\frac13)$ and 
$\hat m=\mbox{diag }(m_u,m_d,m_s)$ are the quarks' electric
charge and mass matrix, respectively.
The result (\ref{first}) is analogous to a similar one obtained in 
\cite{GrNoRo} for the long-distance contribution to the decay 
$B\to \pi e^+e^-$.

The remaining contribution can be expressed in terms of the
two form factors $f_{V,A}(E_\gamma)$ parametrizing the radiative 
leptonic decay $B^-\to\gamma e\bar \nu$
\bea\label{formfs}
& &\frac{1}{\sqrt{4\pi\alpha}}
\langle\gamma(q,\epsilon_\lambda) 
|\bar q\gamma_\mu (1-\gamma_5) b| \bar B(v)\rangle =\\
& &\qquad i\varepsilon(\mu,\epsilon_\lambda^*,v,q) 
f_V(E_\gamma) +
[\epsilon_{\lambda\mu}^* (v\cdot q) - q_\mu 
(\epsilon_\lambda^*\cdot v)]f_A(E_\gamma)
- \frac{1}{E_\gamma}
(Q_q-Q_b)f_B m_B (v\cdot \epsilon^*_\lambda) v_\mu\,,\nonumber
\eea
where $v$ denotes the $B$ meson velocity ($p_B=m_B v$).
The last term is present only for charged $B$ mesons, and is
required for the gauge invariance of the complete decay amplitude,
including also the lepton part.
Although it is not relevant for real photon processes,
it does contribute to decays involving virtual
photons, which can probe the $\mu=0$ component of the electromagnetic
current. A detailed derivation of this term is presented in the 
Appendix.

Combining the two terms in (\ref{11}) one finds the following
result for the weak annihilation helicity amplitudes $A_\lambda$
\bea\label{ALR}
A_{L,R} = -\frac{G_F}{\sqrt2} a_1 e m_{\rho^-} f_{\rho^-} 
\left[f_B + E_\gamma(f_A^{(B^-)}(E_\gamma) \mp
f_V^{(B^-)}(E_\gamma))\right]\,.
\eea
Eventually the form factors $f_{V,A}(E_\gamma)$ will be extracted from
the doubly differential spectrum $d^2\Gamma/dE_e dE_\gamma$ in the 
radiative leptonic decay $B^\pm\to\gamma e\nu$, which will allow a
model-independent calculation of the weak annihilation amplitudes
$A_{L,R}$. A considerable simplification can be achieved if the $WA$
amplitudes are expanded in powers of $\Lambda/E_\gamma$ using the
methods of perturbative QCD for exclusive processes \cite{BrLe}.
It has been shown in \cite{KPY} that the leading terms in the 
expansion (of order $O(1/E_\gamma)$) of $f_V(E_\gamma)$ and 
$f_A(E_\gamma)$
are related, and can be expressed in terms of the valence light-cone
wave function of the $B$ meson as
\bea\label{fVA}
f_V^{(B^\pm)}(E_\gamma)= \pm f_A^{(B^\pm)}(E_\gamma) =
\frac{f_B m_B}{2E_\gamma} \left( Q_u R - \frac{Q_b}{m_b}\right)
+ {\cal O}\left(\frac{\Lambda^2}{E_\gamma^2}\right)\,.
\eea
$R$ is a hadronic parameter given by an integral over the
$B$ meson light-cone wave function $R=\int_0^\infty 
dk_+\psi_B(k_+)/k_+$ (with the normalization 
$\int_0^\infty dk_+\psi_B(k_+)=1$).
A similar result is obtained in the quark
model, with the identification $R\to 1/m_u$, with $m_u\simeq 350$
MeV the constituent light quark mass \cite{AES,ABS,HYCeff}.
The relation (\ref{fVA}) among $f_{V,A}(E_\gamma)$ implies that a 
measurement of the photon spectrum in $B^\pm\to\gamma e\nu$ should
be sufficient for their extraction
(without any knowledge of the $B$ meson light-cone wavefunction).

For the moment, in the absence of such data, the parameter $R$ can
be estimated using model wavefunctions, which results in typical 
values around $R=2.5\pm 0.5$ GeV$^{-1}$ \cite{KPY}. We will use this
central value in our estimates below. 

It is possible to give a model-independent lower bound for
the $R$ parameter, in terms of the $B$ meson binding energy 
$\bar\Lambda=m_B-m_b$. At tree-level this bound reads 
$R \geq 3/(4\bar\Lambda)$ \cite{KPY}. This 
can be used together with (\ref{ALR}) and (\ref{fVA}) to derive a 
lower bound on the magnitude of the $A_L$ amplitude (corresponding to 
$B^-\to\rho^-\gamma$ decays)
\bea\label{AL}
A_L \geq \sqrt{2} G_F a_1 em_{\rho^-} f_{\rho^-} 
\frac{3Q_u f_B m_B}{8\bar\Lambda}
= (0.54\times 10^{-6})
\left(\frac{a_1}{1.0}\right)
\left(\frac{f_B\mbox{ (MeV)}}{175}\right)
\left(\frac{350}{\bar\Lambda\mbox{ (MeV)}}\right)
\mbox{ GeV}\,.
\eea
We used in this estimate $f_{\rho^-}=216$ MeV, as determined from the 
leptonic decay width of the $\rho^0$ meson.

To leading order in $1/E_\gamma$, the right-handed amplitude $A_R$
receives contributions only from the first term in (\ref{ALR})
\bea
A_R = -\frac{G_F}{\sqrt2} a_1 em_{\rho^-} f_{\rho^-} f_B = 
-(0.07\times 10^{-6})
\mbox{ GeV}\,,
\eea
where we used the same values for the hadronic parameters as above.
Comparing with (\ref{AL}) one can see that the left/right-handed WA
amplitude ratio is enhanced by a factor of more than 8.

The $W$-exchange amplitude $E$ is given by an expression analogous 
to (\ref{11}), with a different phenomenological factorization
coefficient $a_2=C_2(m_b) + C_1(m_b)/N_c = 0.17$ \cite{BuBuLa}. 
The result
corresponding to the decay $\bar B^0\to\rho^0\gamma$ is written as
\bea
E_\lambda = G_F a_2 \left\{
-f_B p_{B\mu} 
\langle\rho^0\gamma_\lambda |(\bar uu)_{V-A}^\mu|0\rangle
+ m_{\rho^0} f_{\rho^0} (\epsilon^*_{\rho})_\mu\langle \gamma_\lambda |
(\bar db)_{V-A}^\mu|\bar B^0\rangle \right\}\,.
\eea
In the chiral limit only the second term contributes, which gives
(with $f_{\rho^0}=\frac{1}{\sqrt2}f_{\rho^\pm}$)
\bea
E_{L,R} = -\frac{G_F}{\sqrt2} a_2 em_{\rho^0} f_{\rho^+} E_\gamma 
(f_A^{(\bar B^0)}(E_\gamma) \mp f_V^{(\bar B^0)}(E_\gamma))\,.
\eea
The form-factors appearing on the RHS are given by an expression
analogous to (\ref{fVA}) (the sign change is due to our phase
convention for the $\bar B^0$ state)
\bea\label{fVA1}
f_V^{(\bar B^0)}(E_\gamma)= - f_A^{(\bar B^0)}(E_\gamma) =
-\frac{f_B m_B}{2E_\gamma} \left( Q_d R - \frac{Q_b}{m_b}\right)
+ {\cal O}\left(\frac{\Lambda^2}{E_\gamma^2}\right)\,.
\eea
To the leading order in $\Lambda/E_\gamma$,
one finds the lower bound
\bea
E_L \geq \sqrt{2} G_F a_2 em_{\rho^0} f_{\rho^+} 
\frac{3|Q_d| f_B m_B}{8\bar\Lambda}
= (0.05\times 10^{-6})
\left(\frac{a_2}{0.175}\right)
\left(\frac{f_B\mbox{ (MeV)}}{175}\right)
\left(\frac{350}{\bar\Lambda\mbox{ (MeV)}}\right)
\mbox{ GeV}\,.
\eea
The $W$-exchange amplitude $E$ is suppressed relative to the WA
amplitude $A$ by color-suppression in the ratio $a_2/a_1$ and by
an additional factor $|Q_d/Q_u|=1/2$.
The right-handed $E_R$ amplitude receives contributions only from
the higher twist terms in (\ref{fVA1}) and is therefore
suppressed relative to the left-handed amplitude $E_R$ by a factor 
of $\Lambda/E_\gamma \simeq 0.15$,
corresponding to a photon energy in $B\to\rho(K^*)\gamma$ 
decays $E_\gamma = 2.6$ GeV. This gives the estimate $E_R\simeq
0.15 E_L > (0.007\times 10^{-6})$ GeV.

The remaining amplitudes $P_{u,c}$ are considerably more difficult
to evaluate. A great deal of effort has been put into attempts
to calculate them, with methods such as simple vector meson dominance
\cite{ABS,VMD,HYC}, dispersion techniques combined with Regge theory 
\cite{DoGoPe} and light-cone QCD sum rules \cite{LD1,LD1.5,LD2}. 
In the following section we will revisit some of these estimates,
putting them into perspective relative to the effects computed
in this section (see Table I).

\begin{center}
\begin{tabular}{|c|c|c|c|c|c|}
\hline
Photon helicity & $|P_{t\lambda}|$ & $|P_{c\lambda}|$ & 
$|P_{u\lambda}|$ & $|A_\lambda|$ & $|E_\lambda|$\\
\hline
\hline
$\lambda=L$ & $1.8$ & $0.16$ & $0.03$ &
  $ 0.6$ & $ 0.05$ \\
$\lambda=R$ & $0$ & $0.04$ & $0.007$ & $0.07$ & $0.007$ \\
\hline
\end{tabular}
\end{center}
\begin{quote} {\bf Table I.}
Estimates of the short-distance and long-distance 
amplitudes in $B\to \rho\gamma$ decays (in units of $10^{-6}$ GeV).
The estimates of the $WA$ and $W$-exchange amplitudes $A_\lambda$ 
and $E_\lambda$ 
used $R=2.5$ GeV$^{-1}$.
\end{quote}

\subsection{Long-distance amplitudes with internal $c-$ and $u$-quark
loops}

The long-distance amplitude $P_{c\lambda}$ induced by the $b\to c\bar 
cq$ part of the weak Hamiltonian (\ref{Hweak}) is usually assumed to 
be dominated by the diagram wherein the photon couples to the charm 
quark loop. In a hadronic language, this contribution arises from
the weak decay of the $B$ meson into $V\psi^{(n)}$ (with $\psi^{(n)}$
a $\bar cc$ state with the quantum numbers of the photon), followed 
by the annihilation of the $\psi^{(n)}$ state
into a photon \cite{VMD,HYC}. For an arbitrary photon virtuality 
$q^2$, this long-distance amplitude can be written as a sum over
intermediate states
\beq\label{sumccbar}
A(\bar B\to V\psi^{(n)}\to V\gamma_\lambda) = 
Q_c e\sum_{n,\varepsilon_n} 
\frac{\langle 0|
\bar c\gamma\cdot \varepsilon_\lambda^* c|\psi^{(n)}(q,\varepsilon_n)
\rangle A(\bar B\to V\psi^{(n)})}
{q^2-M_{n}^2+iM_n \Gamma_n}\,.
\eeq
The lowest-lying state contributing to the sum is the $J/\psi$,
with a mass of 3.097 GeV. Therefore, for a real photon, the
$J/\psi$ propagator denominator is large, such that its contribution 
can be expected to be strongly suppressed. These
heuristic arguments are 
supported by an explicit QCD sum rule calculation \cite{LD2}, where 
the charmed penguin amplitude $P_{c\lambda}$ is found to be less than
5\% of the dominant short-distance amplitude $P_{tL}$. This result 
is confirmed by existing experimental data, as discussed below. 

In the following we will substantiate these qualitative arguments 
with an estimate of the sum (\ref{sumccbar}), truncating it to a
few lowest states. For definiteness we 
consider the decay $B^-\to \rho^-\gamma$. The weak decay amplitude
in (\ref{sumccbar}) can be estimated using factorization, with the
result
\bea\label{Aweak}
A(B^-\to \rho^-\psi^{(n)}) &=& \frac{G_F}{\sqrt2}\lambda_c^{(d)}
a_2 f_{\psi^{(n)}}m_{\psi^{(n)}}
\left\{
\frac{2V(m^2_{\psi^{(n)}})}{m_B+m_\rho}\,
i\varepsilon(\epsilon_\psi^*,p_B,p_\rho,\epsilon_\rho^*)\right.\\
& &\left. - (m_B+m_\rho) A_1(m^2_{\psi^{(n)}})
(\epsilon_\psi^*\cdot\epsilon_\rho^*)
+ \frac{2A_2(m^2_{\psi^{(n)}})}{m_B+m_\rho}( \epsilon_\rho^*\cdot q)
(\epsilon_\psi^*\cdot p_B)
\right\}\nonumber
\eea
The form-factors appearing here are defined in the usual way
(with $q=p'-p$)
\bea
& &\langle \rho^-(p',\epsilon)|\bar u\gamma_\mu b|B^-(p)\rangle =
\frac{2V(q^2)}{m_B+m_\rho}i\varepsilon(\mu,p,p',\epsilon^*)\\
& &\langle \rho^-(p',\epsilon)|\bar u\gamma_\mu\gamma_5 b|
B^-(p)\rangle =
2m_\rho A_0(q^2)\frac{\epsilon^*\cdot q}{q^2}q_\mu +
(m_B+m_\rho)A_1(q^2)\left(\epsilon^*_\mu-
\frac{\epsilon^*\cdot q}{q^2}q_\mu\right)\\ 
& &\qquad\qquad- A_2(q^2)\frac{\epsilon^*\cdot q}{m_B+m_\rho}
\left(p_\mu + p'_\mu -
\frac{m_B^2-m_\rho^2}{q^2}q_\mu\right)\,,\nonumber
\eea
and the decay constant of the $\psi$ state is given by
$\langle 0|\bar c\gamma_\mu c |\psi(q,\epsilon)\rangle =
m_\psi f_\psi \epsilon_\mu$. Inserting (\ref{Aweak}) into the
sum (\ref{sumccbar}) and performing the sum over the $\psi^{(n)}$
polarization one finds the following result for the long-distance 
charmed penguin contribution to the $B^-\to \rho^-\gamma$ amplitude
\bea
P_{c\lambda}(B^-\to \rho^-\gamma_\lambda) &=& 
Q_c e \frac{G_F}{\sqrt2} 
a_2 \frac{2m_B}{m_B+m_\rho} \sum_n f^2_{\psi^{(n)}} 
\\
& &\times 
\left\{ V(0)\, i\varepsilon(q,\epsilon_\lambda^*,\epsilon_\rho^*,v) - 
A_2(0) [(v\cdot q)(\epsilon_\rho^*\cdot\epsilon_\lambda^*) - 
(v\cdot\epsilon_\lambda^*)(q\cdot \epsilon_\rho^*)]\right\}\,.\nonumber
\eea

To put the result into this (explicitly gauge-invariant) form,
one must assume the following relation between the form factors
$A_1$ and $A_2$ at $q^2=0$ \cite{HYC}
\bea\label{A1A2}
A_1(0) = A_2(0)\frac{m_B-m_\rho}{m_B+m_\rho}\,.
\eea
While the approximate numerical values of these form factors 
$V^{B\rho}(0)=0.34, A_1^{B\rho}(0)=0.26, A_2^{B\rho}(0)=0.22$ \cite{BB}
do not 
satisfy exactly the relation (\ref{A1A2}), it is worth noting that it
does hold true to leading order in a large energy expansion in powers 
of $1/E_\rho$.
In \cite{Charles}, the form-factors for the decay of a $B$ meson into 
a light pseudoscalar (vector) meson have been analyzed using an
effective theory for the final state quarks \cite{DuGr}. To leading 
twist, all these form factors can be expressed in terms of two 
universal 
functions $\zeta_\perp(E_\rho)$ and $\zeta_\parallel(E_\rho)$
\bea\label{rel1}
V(q^2) &=& \left(1 + \frac{m_\rho}{m_B}\right)\zeta_\perp(E_\rho)\\
\label{rel2}
A_1(q^2) &=& \frac{2E_\rho}{m_B+m_\rho}\zeta_\perp(E_\rho)\\
\label{rel3}
A_2(q^2) &=& \left(1 + \frac{m_\rho}{m_B}\right)\left[
\zeta_\perp(E_\rho) - \frac{m_\rho}{E_\rho}\zeta_\parallel(E_\rho)
\right]\,.
\eea
The leading twist form factors satisfy certain relations, which are 
very similar (but not completely identical) to relations previously 
derived in the quark model \cite{Stech}. This is not completely
unexpected, considering that to leading twist only the meson's valence 
degrees of freedom are relevant, which esentially coincides
with the quark model picture.

Using the relations (\ref{rel2}), (\ref{rel3}) one can see that the
condition (\ref{A1A2}) is indeed satisfied, up to corrections of
higher order in $m_\rho/E_\rho$, proportional to $\zeta_\parallel$.
The relations (\ref{rel1}), (\ref{rel3}) imply also that only
left-handed photons are emitted to leading twist. The relevant
helicity amplitudes are given by
\bea
P_{cL,R} &=& Q_c e \frac{G_F}{\sqrt2}
a_2 \frac{2m_B}{m_B+m_\rho} E_\gamma (V(0)\pm A_2(0))
\sum_n f^2_{\psi^{(n)}}\,.
\eea
Using the realistic values for form-factors quoted above, one finds that
the left/right-handed helicity amplitudes ratio is enhanced by about
a factor of 5. Absolute values for the $|P_{c\lambda}|$ helicity 
amplitudes obtained by keeping only the lowest two states 
in (\ref{sumccbar}) are quoted in Table I. The corresponding decay
constants can be extracted from their leptonic widths and are given by
$f_{\psi^{(1)}}=395$ MeV, $f_{\psi^{(2)}}=293$ MeV \cite{HYC}.

A similar analysis can be performed for the long-distance amplitude
$P_{u\lambda}$ describing quark diagrams with $u$-quark loops, which 
is given by a sum analogous to (\ref{sumccbar})
\beq\label{sumuubar}
A(\bar B\to VV'\to V\gamma_\lambda) = e\sum_{V'=\rho^0, \omega,\dots} 
\frac{\langle 0|
j_{\rm e.m.}\cdot \varepsilon_\lambda^* |V'(q,\varepsilon)
\rangle A_C(\bar B\to VV')}{q^2-M_{V'}^2+iM_{V'} \Gamma_{V'}}\,.
\eeq
The weak amplitudes $A(\bar B\to VV')$ receive, in general, both
color-allowed and color-suppressed contributions (in contrast to
the decays $B\to V\psi$ which are purely color-suppressed). 
For example, in $B^-\to\rho^-\gamma$ decay, the color-allowed 
component gives a WA-type contribution to the radiative decay. To 
avoid double counting of amplitudes, we will keep therefore only the 
color-suppressed component of the weak amplitude in (\ref{sumuubar}) 
(symbolized by the subscript $C$). Our analysis is different in this
respect from other VMD estimates of this amplitude \cite{HYC}.

Keeping only the contributions of the lowest two states $V'=\rho^0,
\omega$, one finds
\bea
P_{uL,R} = \frac{G_F}{\sqrt2} e a_2 f_{\rho_0} m_{\rho_0}
\frac{2m_B}{m_B+m_\rho} E_\gamma \left(
\frac{f_{\rho_0}}{m_{\rho}} + \frac{f_{\omega}}{m_{\omega}}
\right)
(V(0)\pm A_2(0))\,.
\eea
In writing this result we used the nonet symmetry relation
$\langle 0|\bar u\gamma_\mu u |\rho^0\rangle = 
\langle 0|\bar u\gamma_\mu u |\omega\rangle$, valid in the large
$N_c$ limit or the quark model. The corresponding
numerical results are shown in Table I.

The enhancement of the left-handed photon amplitudes relative to the
right-handed ones in the long-distance contributions we found is 
somewhat surprising. Calculations of these amplitudes using different 
methods such as QCD sum rules \cite{LD1,LD1.5,LD2} and perturbative 
QCD \cite{pQCD} find a similar result. 
Such an enhancement does not necessarily hold in the presence of new
physics. In certain extensions of the Standard Model, such as the
left-right symmetric model, the right-handed helicity amplitude can
become large (proportional to the virtual heavy fermion mass)
\cite{ChoMi}. Such effects lead to observable consequences such as
significant mixing-induced CP asymmetries in weak radiative 
$B\to V\gamma$ decays \cite{AtGrSo}.

To summarize the results of this section, the rare radiative decays
are dominated by the amplitudes corresponding to left-handed photons.
The dominant long-distance contribution to the $B^-\to\rho^-\gamma$ 
decay comes from the $WA$ mechanism; the corresponding amplitude
$A_\lambda$
is calculable in a model-independent way from experimental data.
The left-handed amplitudes satisfy an approximate hierarchy of sizes 
\bea
E \simeq P_{u} \simeq 0.4P_{c}\,,\quad
P_{c} \simeq 0.2 A\,,\quad A\simeq 0.3 P_{t}\,.
\eea
The relative magnitude of the components with different CKM 
coefficients can be obtained by combining these estimates for
the amplitudes with the corresponding CKM factors.





\subsection{Nonfactorizable corrections}

The leading corrections to the factorization result (\ref{11}) for the
WA amplitudes come, in perturbation theory, from the graphs in Fig.~2,
with one gluon exchanged between the $B$ and $\rho$ quarks. 

Let us take for definiteness the final vector meson to be moving
along the positive $z$ axis. Momentum conservation
$m_B v = q + p$ is expressed in terms of light-cone components as
$p_+ = m_B$ and  $m_B = 2E_\gamma + p_-$ (we define the 
light-cone coordinates as $p_\pm = p^0\pm p^3$). Then to leading order
$O(1)$ in an expansion in powers of $\Lambda/p_+=\Lambda/m_B$, the
$\rho$ meson can be represented by its valence component 
$|q(p_1)\bar q(p_2)\rangle$
with parallel momenta $p_1 = (yp_+, 0_-, 0_\perp)$ and 
$p_2 = ((1-y)p_+, 0_-, 0_\perp)$. In a tensor language this corresponds
to the $\rho$ wave function
\beq
(\Psi_\rho)_{\alpha\beta} = \frac{\sqrt{N_c}}{8\sqrt2} 
p_+(\gamma_-\epsslash)_{\alpha\beta} \phi_\perp(y)
\eeq
where $\phi_\perp(y)$ is the twist-2 chiral-odd structure function
as appropriate for a transversely polarized $\rho$ meson. It 
satisfies the normalization condition
\beq
\int_0^1 dy \phi_\perp(y) = \sqrt{\frac{2}{N_c}}f_\rho^T\,,
\eeq
where the decay constant $f_\rho^T = (160\pm 10)$ MeV \cite{rho}
is defined as
\beq
\langle 0|\bar u\sigma_{\mu\nu} d|\rho^-(p,\varepsilon)\rangle
 = if_\rho^T (\varepsilon_\mu p_\nu - \varepsilon_\nu p_\mu)\,.
\eeq
The $B$ meson is represented by the tensor wavefunction \cite{KPY}
\bea
(\Psi_B)_{\alpha\beta} =
\frac{\sqrt{N_c}}{\sqrt2 k_+}\psi_B(k_+,\vec k_\perp)
\left\{ (k_+ + \vec\alpha_\perp\cdot\vec k_\perp)
\Lambda_+ \frac{1+\vslash}{2}\gamma_5\right\}_{\alpha\beta}\,,
\quad \Lambda_+ = \frac{\gamma_-\gamma_+}{4}\,.
\eea
In contrast to the light mesons $\pi$ or $\rho$, the leading twist
description of a heavy meson involves in general (beyond tree level) 
also the transverse motion of the heavy quark.

Each of the four diagrams in Fig.~2a-d contains IR divergences;
however, an explicit calculation shows that they cancel in the sum.
Therefore, (at least to one-loop order) the leading-twist 
nonfactorizable contribution is computable in perturbation theory as 
a convolution of the $B$ and $\rho$ light-cone wave functions with a 
IR-finite hard scattering amplitude $T_H$ 
\bea
\delta_{n.f.} A_L = \int_0^\infty dk_+ \int_0^1 dy \phi_\perp(y,\mu)
\psi_B(k_+,\mu) T_H(k_+,y,\mu) + \mbox{higher twist}\,.
\eea
(This is analogous to a result recently found in \cite{BBNS} for
nonleptonic decays $B\to\pi\pi$.)
However, it is easy to see that the hard-scattering amplitude
vanishes in the limit of massless final quarks $T_H = O(m_{u,d})$.
Technically, in this limit, the $\rho$ side of the matrix
element contains only traces of an odd number of $\gamma$ matrices,
which vanish.

None of these conclusions depends on the fact that the photon is
on-shell $q^2=0$, such that a similar result can be derived for 
$B\to \rho_\perp e^+e^-$ decays, mediated by a virtual photon. 
However, a nontrivial nonfactorizable correction is obtained for a
longitudinally polarized $\rho$ or a pion in the final state.

In conclusion, the one-loop nonfactorizable correction 
vanishes identically in the chiral limit and to leading-twist order.
Therefore the factorized result (\ref{11}) can be expected to
give a very good approximation to the weak annihilation amplitude $A$.

\section{Applications}
\subsection{Implications from experimental data}

The CLEO Collaboration  \cite{CLEOexcl} recently reported the 
branching ratios for the $B\to K^*\gamma$ exclusive modes
\bea\label{exp1}
{\cal B}(B^\pm \to K^{*\pm}\gamma) &=& (3.76^{+0.89}_{-0.83}\pm
0.28)\times 10^{-5}\\\label{exp2}
{\cal B}(B^0 \to K^{*0}\gamma) &=& (4.55^{+0.72}_{-0.68}\pm
0.34)\times 10^{-5}\,.
\eea
These results give the following ratio for the charge-averaged
radiative decay widths
\bea\label{ratioexp}
\frac{\Gamma(B^0 \to K^{*0}\gamma)}{\Gamma(B^\pm\to K^{*\pm}\gamma)}=
\frac{\tau_{B^\pm}}{\tau_{B^0}}
\frac{{\cal B}(B^0 \to K^{*0}\gamma)}
{{\cal B}(B^\pm \to K^{*\pm}\gamma)}
 = 1.29\pm 0.55\,.
\eea
We used here the lifetime ratio of charged and neutral $B$
mesons $\tau(B^\pm)/\tau(B^0) = 1.066\pm 0.024$ \cite{BLEP}.

Isospin symmetry constrains the short-distance amplitude $P_{tL}$ to
be the same in both decays (\ref{exp1}), (\ref{exp2}). Therefore the
deviation of the ratio (\ref{ratioexp}) from 1 can
only come from the long-distance contribution $M^{(2)}-P_c^{(2)}$
arising from the photon coupling to the spectator quark.
Using the unitarity of the CKM matrix one can rewrite $\lambda_c^{(s)}
= - \lambda_u^{(s)} - \lambda_t^{(s)}$ in (\ref{6}) and (\ref{8}).
Furthermore, neglecting the small CKM factor $\lambda_u^{(s)}$, these
amplitudes can be written to a good approximation as
\bea
A(B^- \to K^{*-}\gamma) &=& \lambda_t^{(s)}\left(
P_{t} + (M^{(1)}-P_c^{(1)}) + \frac23(M^{(2)}-P_c^{(2)})\right)\\
A(\bar B^0 \to \bar K^{*0}\gamma) &=& \lambda_t^{(s)}\left(
P_{t} + (M^{(1)}-P_c^{(1)}) - \frac13(M^{(2)}-P_c^{(2)})\right)\,.
\eea

The short-distance amplitude $P_{tL}$ alone would give the following
branching ratios
\bea\label{th1}
{\cal B}_{\rm s.d.}(B^\pm \to K^{*\pm}\gamma) &=& 
4.6\times 10^{-5}\,,\qquad
{\cal B}_{\rm s.d.}(B^0 \to K^{*0}\gamma) = 4.5\times 10^{-5}\,,
\eea
where we used the estimate for $P_{tL}$ shown in Table I together with
$\lambda_t^{(s)}=0.039$. The $B$ meson lifetimes were taken
as $\tau_{B^\pm}
= 1.64\pm 0.02$ ps, $\tau_{B^0} = 1.55\pm 0.03$ ps \cite{BLEP}.
Assuming that the central values of the experimental numbers 
(\ref{exp1}), (\ref{exp2}) will be confirmed by future, more precise
determinations, one is led to conclude therefore
that the long-distance effects 
interfere among themselves destructively in $B^0$ decays, whereas in 
$B^{\pm}$ they could be significant, and contribute up to 10\% of the 
total amplitude. 
This agrees with the theoretical estimates of \cite{pQCD} and disfavor
a non-SM explanation of the isospin breaking in the ratio 
(\ref{ratioexp}) based on an enhanced gluonic penguin as proposed in 
\cite{Petrov}.

In the SU(3) limit the long-distance contributions to $B^0$ and $B_s$
decays are the same (see (\ref{7})-(\ref{12})). Therefore one can use 
the observed $B^0\to K^{*0}\gamma$ branching ratio to make 
predictions for other
strangeness-changing radiative decays of these mesons. Neglecting the 
OZI-suppressed amplitudes $S_{c\lambda}$ and $PA_{c\lambda}$,
one obtains the following prediction (together
with $\tau_{B_s}=1.46\pm 0.06$ ps \cite{BLEP})
\bea
{\cal B}(B_s\to \phi\gamma) &=& 
\frac{\tau_{B^0}}{\tau_{B_s}}
{\cal B}(B^0\to K^{*0}\gamma) \simeq 4.8\times 10^{-5}\,.
\eea
Using the estimate for $E_L$ in Table I, and neglecting again the
contributions of the OZI-suppressed annihilation penguin $PA$ and 
SU(3) singlet $S$ amplitudes, one
can predict very small rates for the $B_s$ modes
(we used here $|V_{ub}|=1.6\times 10^{-2}$)
\bea\label{Bsrho}
{\cal B}(B_s\to \rho^0\gamma) \simeq
{\cal B}(B_s\to \omega\gamma) \simeq 0.3\times 10^{-8}\,.
\eea

\subsection{Determining $|V_{td}|$}

The main interest in observing the exclusive charmless radiative
decay $B^\pm\to \rho^\pm\gamma$ is connected with the possibility
of determining the CKM matrix element $|V_{td}|$.
The amplitude for this decay is related by $U$-spin symmetry to
that for the $b\to s\gamma$ mode $B^\pm\to K^{*\pm}\gamma$
\bea
A(B^-\to \rho^-\gamma_L) &=& 
\lambda_u^{(d)}a_L + \lambda_t^{(d)}p_L\,,
\qquad\qquad A(B^-\to K^{*-}\gamma_L) = 
\lambda_u^{(s)}a'_L + \lambda_t^{(s)}p'_L\,.
\eea
In the SU(3) symmetry limit $a_L=a_L'=A_L-P_{cL}, p_L=p_L'=
P_{tL}+M_L-P_{cL}$.
In the absence of the $a, a'$ terms, the ratio of the two amplitudes
is equal to $|\lambda_t^{(d)}/\lambda_t^{(s)}|=|V_{td}/V_{ts}|$
(up to SU(3) breaking corrections), which offers a way for 
determining $|V_{td}|$.

In the general case, however, the $(\rho^-\gamma_L)$ amplitude
can be written as
\bea
A(B^-\to \rho^-\gamma_L) &=& \lambda_t^{(d)}p(1 - 
\frac{|\lambda_u^{(d)}|}{|\lambda_t^{(d)}|} e^{i\alpha}
\varepsilon_A e^{i\phi_A})\,,
\eea
with $\varepsilon_A e^{i\phi_A} \equiv a/p \simeq A_L/P_{tL}$ 
(where, following the estimates of Sec.~II we neglected 
the charmed penguin $P_{cL}$ and gluonic penguin $M_L$ amplitudes
relative to the short-distance amplitude $P_{tL}$ and the $WA$ 
amplitude $A_L$). The
estimates of the preceding section give $\varepsilon_A\simeq 0.3$,
and $\phi_A=0$ to leading twist.
Furthermore, global analyses of the unitarity triangle suggest that
the ratio of CKM factors appearing in this amplitude is subunitary
\cite{AliLo}
\beq
\frac{|\lambda_u^{(d)}|}{|\lambda_t^{(d)}|} \simeq
\left|\frac{V_{ub}}{V_{cb}}\right|
\cdot \left|\frac{V_{ts}}{V_{td}}\right| \simeq 0.4\,.
\eeq

The estimates of Sec.~II.B show that the total rate for
$B^-$ ($B^+$) radiative decay is dominated by left-hand 
(right-hand) polarized photons. Neglecting the small right-handed
(left-handed) component, which gives only a second order contribution
to the ratio, one finds for the ratio of the charge-averaged rates
\bea\label{ratioVtd}
& &
\frac{{\cal B}(B^-\to\rho^-\gamma) + {\cal B}(B^+\to\rho^+\gamma)}
{{\cal B}(B^-\to K^{*-}\gamma) + {\cal B}(B^+\to K^{*+}\gamma)}\\
& &\quad = \left(\frac{|p|}{|p'|}\cdot \frac{|V_{td}|}{|V_{ts}|}\right)^2
\left\{1 - \frac{|\lambda_u^{(d)}|}{|\lambda_t^{(d)}|}\varepsilon_A
\cos\alpha\cos\phi_A + \left(\frac{
|\lambda_u^{(d)}|}{|\lambda_t^{(d)}|}\varepsilon_A\right)^2
\right\}\,.\nonumber
\eea
The factor in the braces is bounded from above and below by using
the inequality
\beq
1 + x + x^2 \leq 1 - x\cos\alpha\cos\phi_A + x^2 \leq 1 - x + x^2\,,
\quad x \equiv \frac{|\lambda_u^{(d)}|}{|\lambda_t^{(d)}|}\varepsilon_A
\simeq 0.12\,.
\eeq 
Assuming, conservatively, $x=0.2$ (which is on the upper side of the 
estimates for this
parameter) gives an uncertainty of about 20\% in the determination
of $|V_{td}|$ arising from the factor in braces in (\ref{ratioVtd}).
The SU(3) breaking factor in the ratio of the penguin amplitudes 
$|p|/|p'|\simeq g_+^{(\rho)}(0)/g_+^{(K^*)}(0) = 0.76\pm 0.22$ is 
known with a uncertainty of about 30\% (we used here the results of 
the LCSR calculation \cite{BB}). Future lattice QCD calculations will
hopefully improve the precision with which this ratio is known.
Adding these errors in quadrature 
one finds a theoretical error in the determination of 
$|V_{td}|$ from the ratio (\ref{ratioVtd})  of about $35\%$.
The leading twist result $\phi_A=0$ implies that the CP asymmetry
in $B^\pm\to\rho^\pm\gamma$ can be expected to be very small 
$(A_{CP}\propto \sin\phi_A)$.

The above estimate for the long-distance contribution $\varepsilon_A$
is in general agreement with the quark model
of \cite{ABS,HYC} and QCD sum rules \cite{LD1,LD1.5} calculations. 
The advantage of our approach will become apparent
once experimental data on radiative leptonic $B$ decays become
available. Such data could be used, as discussed in Sec.~II, to
actually determine the WA amplitude $A_L$, and thereby the
long-distance contamination $\varepsilon_A$.

\section{Conclusions}

Present experimental data from CLEO on exclusive radiative weak 
decays  \cite{CLEOexcl}
are sufficiently precise such that the long-distance effects are
beginning to be observable. A good control over the magnitude of 
these effects is 
important for an assessment of the uncertainty they induce into 
the extraction of $|V_{td}|$ by combining $B^-\to\rho^-\gamma$
and $B^-\to K^{*-}\gamma$ decays \cite{ABS,HYC}. Unfortunately, 
these effects have proved notoriously difficult to treat in any 
systematic way.

In the present paper we focus on a certain class of long-distance
contributions, arising from weak annihilation and $W$-exchange
quark diagram topologies. We argue that the former gives the
dominant long-distance correction to the amplitude for 
$B^-\to\rho^-\gamma$ decays. These corrections can be computed 
reliably using factorization, in terms of form-factors observable 
in radiative leptonic decays $B^+\to\gamma e^+\nu$. The
nonfactorizable corrections are shown to be very small, as they
appear only at nonleading twist. A similar method can be used to
compute the $W$-exchange-type long-distance contribution, relevant
for the weak radiative $B_d$ and $B_s$ decays.

Furthermore, to the leading order of an expansion in powers of 
$\Lambda/E_\gamma$, one can show that the coupling of left-handed
photons in the long-distance amplitude is greatly enhanced relative 
to that of right-handed photons, just as in the short-distance part
of the amplitude.
Such an enhancement had been previously observed in QCD sum rule 
calculations of the long-distance amplitude \cite{LD1}; our
approach clarifies the theoretical limit in which this enhancement
holds and quantifies the magnitude of the corrections to it.

Our results should allow one to reduce the model-dependence of 
the leading long-distance effects in $B^-\to\rho^-\gamma$ decays,
and achieve a better control over the theoretical error in the
corresponding determination of $|V_{td}|$. 

\acknowledgements

D. P. would like to thank Jon Rosner for discussions of the 
quark diagram approach and Vivek Sharma for comments about the 
experimental feasibility of some of the methods discussed here. 
We are grateful to Ahmed Ali and Hai-Yang Cheng for comments on 
the manuscript and to Detlef Nolte for many discussions.
This work has been supported by the National Science Foundation.

\appendix
\newpage
\section{Ward identities for long-distance matrix elements}

We present in this Appendix a set of constraints
on certain long-distance contributions to weak radiative decays
of $B$ mesons. These constraints follow from a Ward identity 
expressing the conservation of the electromagnetic current. Let us 
consider the
following matrix element of a local operator ${\cal O}$, to first
order in electromagnetism and to all orders in the strong coupling
\bea\label{A1}
\langle \gamma(q,\epsilon) f|{\cal O}(0)|B(v)\rangle =
-ie\epsilon_\mu^*\int \mbox{d}^4x e^{iq\cdot x}
\langle f|\mbox{T} j^{\rm e.m.}_\mu(x)\, {\cal O}(0)|B(v)\rangle\,,
\eea
for any hadronic final state $f$. The operator ${\cal O}$ can be
any quark bilinear including gluon fields, or even a four-quark
operator. The electromagnetic current includes contributions from
both the light and heavy quarks $j^{\rm e.m.}_\mu = \bar q\hat {\cal Q}
\gamma_\mu q + \frac23 \bar c\gamma_\mu c - \frac13\bar b\gamma_\mu
b$. A relation similar to (\ref{A1}) can be written for the matrix 
element with the photon replaced with a dilepton pair in the final 
state, coupling through a virtual photon to the hadronic system.

The conservation of the electromagnetic current implies, in the standard
way, a Ward identity for the matrix element of the time-ordered  
product in (\ref{A1})
\bea\label{A2}
-iq_\mu \int \mbox{d}^4x e^{iq\cdot x}
\langle f|\mbox{T} j^{\rm e.m.}_\mu(x)\, {\cal O}(0)|B(v)\rangle = 
\int \mbox{d}^3x e^{-i\vec q\cdot \vec x}
\langle f|[j^{\rm e.m.}_0(\vec x)\,, {\cal O}(\vec 0)]|B(v)\rangle\,.
\eea
The commutator on the RHS is nonvanishing only if the
operator ${\cal O}$ carries an electric charge, as in the case 
of decays induced by the weak charged current such as
$B^+\to \gamma e^+\nu$, 
$B\to \pi(\rho)\gamma e^+\nu$ or $\bar B\to D^{(*)}\gamma e^+\nu$.
In the following we analyze the simplest such case,
the radiative leptonic decays of a $B$ meson, for which ${\cal O} = 
\bar b\gamma_\nu\gamma_5 q$ and $|f\rangle = |0\rangle$.

The equal-time commutator on the RHS of (\ref{A2}) can be computed
explicitly, with the result
\bea\label{A3}
-iq_\mu \int \mbox{d}^4x e^{iq\cdot x}
\langle 0|\mbox{T} j^{\rm e.m.}_\mu(x)\, (\bar
b\gamma_\nu\gamma_5 q)(0)|B(v)\rangle &=& (Q_b - Q_q)
\langle 0|\bar b\gamma_\nu\gamma_5 q|B(v)\rangle\\
&=& (Q_b - Q_q) f_B m_B v_\mu\,.\nonumber
\eea
The most general parametrization of the matrix element on the LHS 
can be written in terms of five form-factors $f_i(q^2,v\cdot q)$
\bea
-i\int \mbox{d}^4x e^{iq\cdot x}
\langle 0|\mbox{T} j^{\rm e.m.}_\mu(x)\, (\bar
b\gamma_\nu\gamma_5 q)(0)|B(v)\rangle = 
f_1 g_{\mu\nu} + f_2 v_\mu v_\nu + f_3 q_\mu q_\nu + 
f_4 q_\mu v_\nu + f_5 v_\mu q_\nu\,.
\eea
The Ward identity (\ref{A3}) implies two constraints on these
form-factors
\bea
(v\cdot q) f_2 + q^2 f_4 = (Q_b-Q_q) f_B m_B\,,\qquad
f_1 + q^2 f_3 + (v\cdot q) f_5 = 0\,.
\eea
For the case of a real photon $q^2=0$, these constraints
fix uniquely the form-factor $f_2(0,v\cdot q)$, and relate
$f_1(0,v\cdot q)$ and $f_5(0,v\cdot q)$. From Eq.~(\ref{A1})
one finds thus the following result for the matrix element of
the axial weak current
\bea
\langle \gamma(q,\epsilon) f|\bar b\gamma_\mu\gamma_5q|B(v)\rangle =
- f_5[(v\cdot q) \epsilon^*_\mu - (v\cdot\epsilon^*) q_\mu]
+ (v\cdot\epsilon^*) v_\mu \frac{1}{v\cdot q}(Q_b-Q_q) f_B m_B\,,
\eea
which is the same as the result presented in text Eq.~(\ref{formfs}),
with the identification $f_5 = f_A$.

\begin{figure}[hhh]
 \begin{center}
 \mbox{\epsfig{file=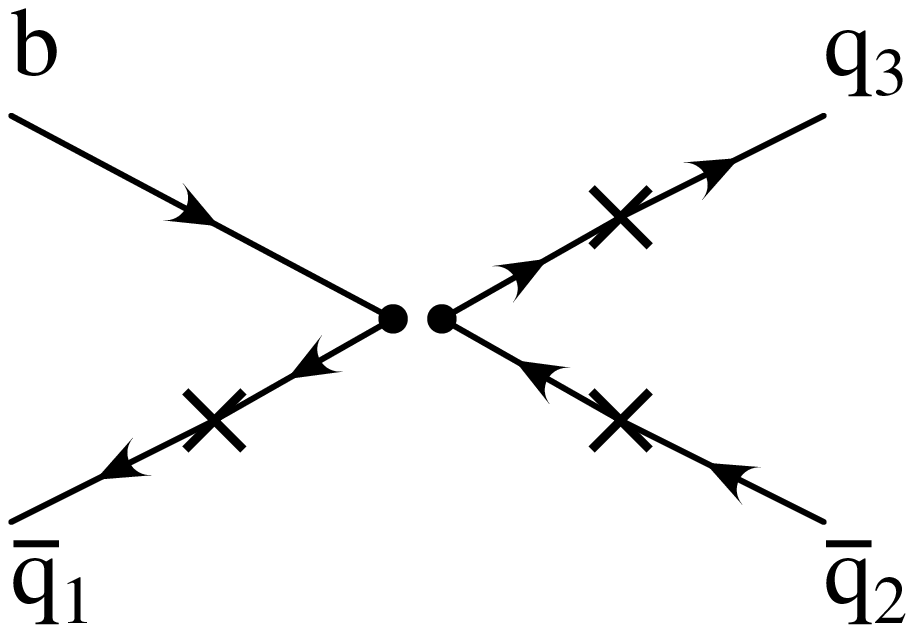,width=4cm}\qquad\qquad
 \epsfig{file=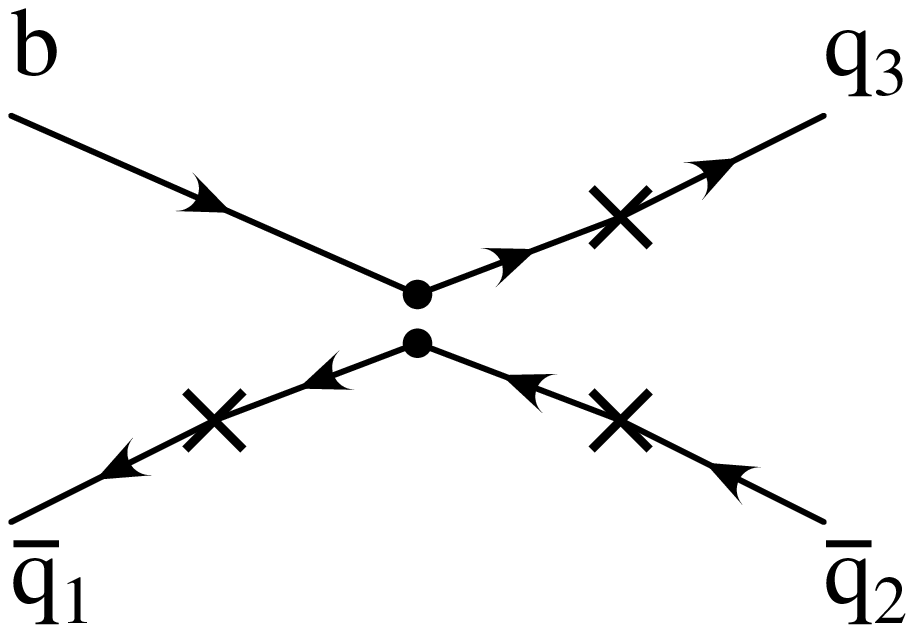,width=4cm}}\\
(a)\hspace*{5cm} (b)\\
 \mbox{\epsfig{file=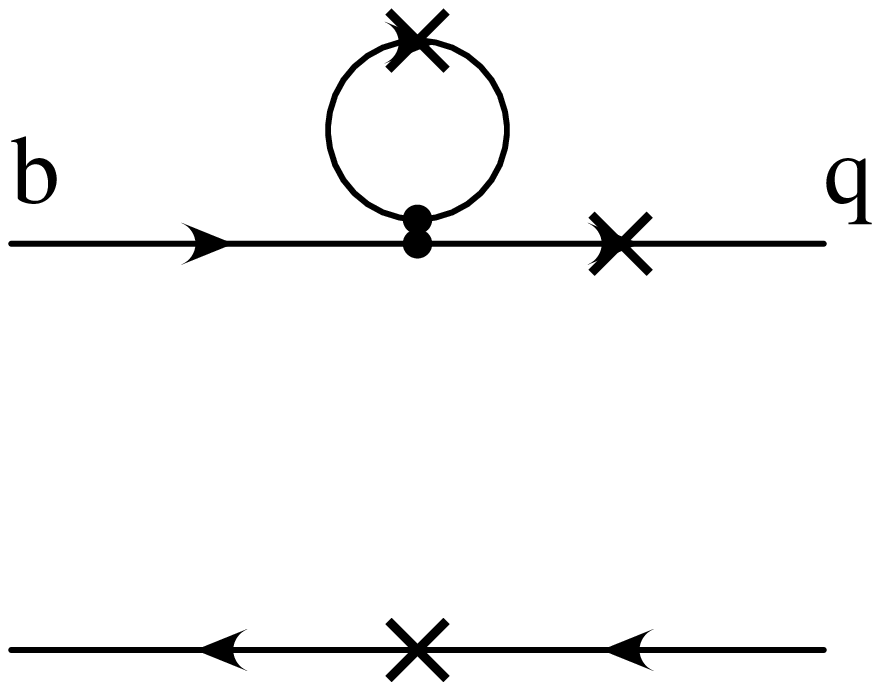,width=4cm}\qquad\qquad
 \epsfig{file=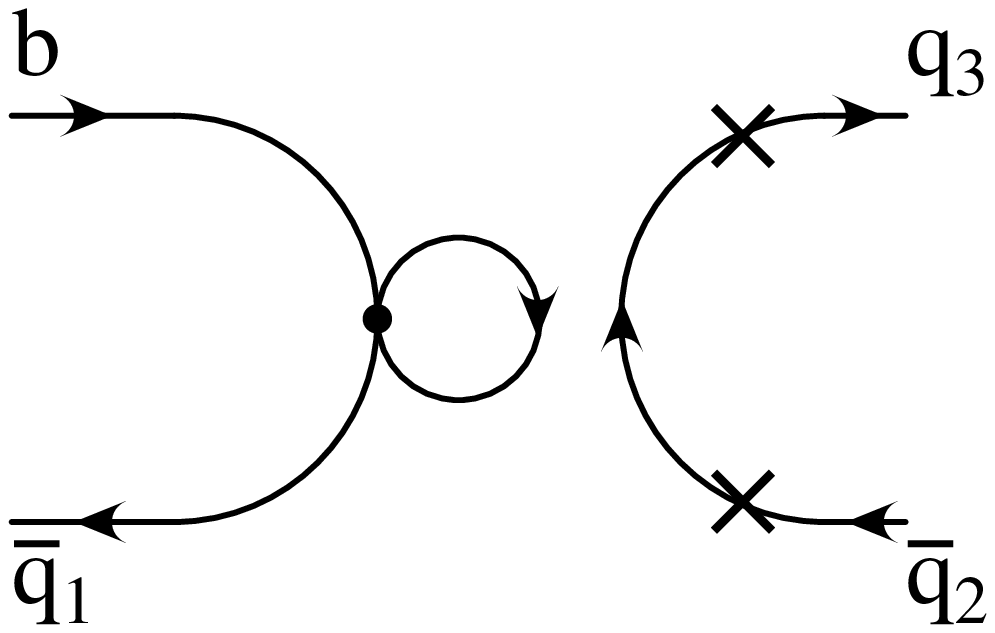,width=4cm}}\\
(c)\hspace*{5cm} (d)\\
 \mbox{\epsfig{file=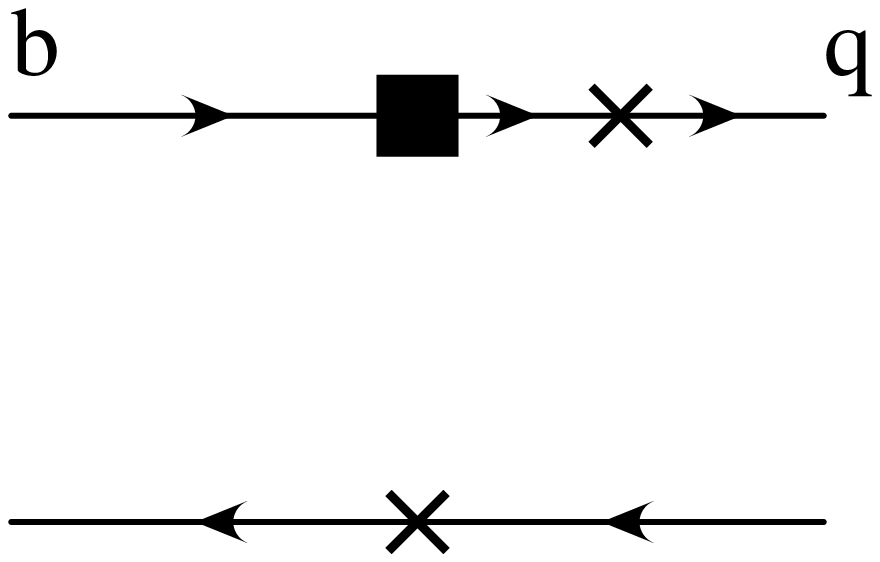,width=4cm}}\\
(e)
 \end{center}
 \caption{
Quark diagrams contributing to $\bar B\to V\gamma$ decays. The cross
marks the attachment of the photon line. a) weak annihilation
amplitudes $A^{(i)}$, with $i=1,2,3$ corresponding to the photon
attaching to each of the three quark lines $\bar q_{1}, \bar q_2, q_3$;
b) $W$-exchange amplitudes $E^{(i)}$, $i=1,2,3$; c) penguin amplitudes
$P_{q'}^{(1)}$ (the photon is attached to the $\bar q=\bar d,\bar s$ 
quark or the quark $q'=u,c$ running in the loop) and $P_{q'}^{(2)}$ 
with the photon attached to the spectator quark; d) annihilation 
penguin amplitudes $PA_{q'}$; e) amplitudes with one insertion of the
gluonic penguin $M^{(1)}$ (photon attaching to the $\bar q$ line) and
$M^{(2)}$ (photon attaches to the spectator quark). The box denotes 
one insertion of the operator $Q_8$.}
\label{fig1}
\end{figure}

\newpage
\begin{figure}[hhh]
 \begin{center}
 \mbox{\epsfig{file=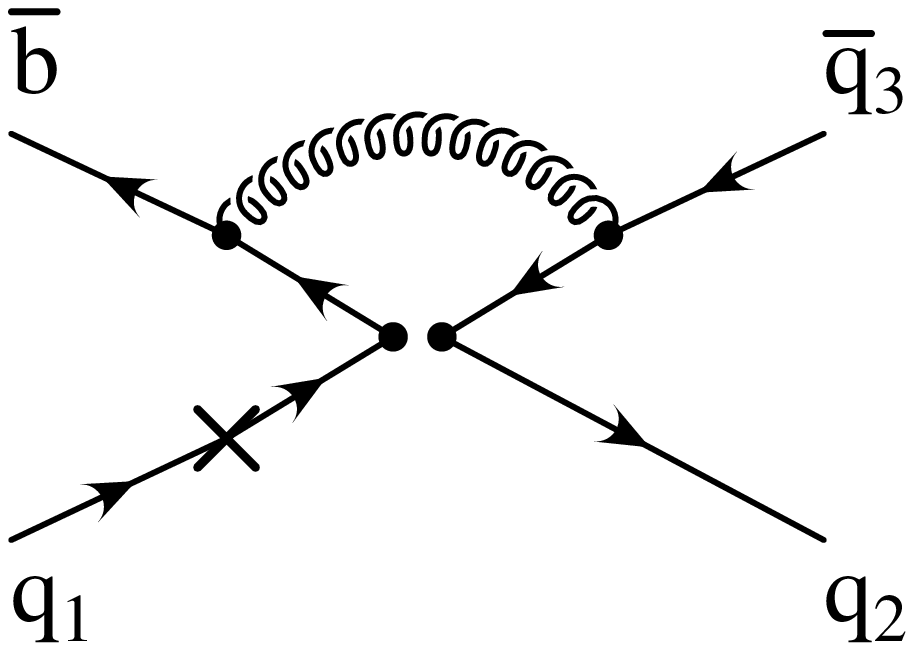,width=4cm}\qquad\qquad
 \epsfig{file=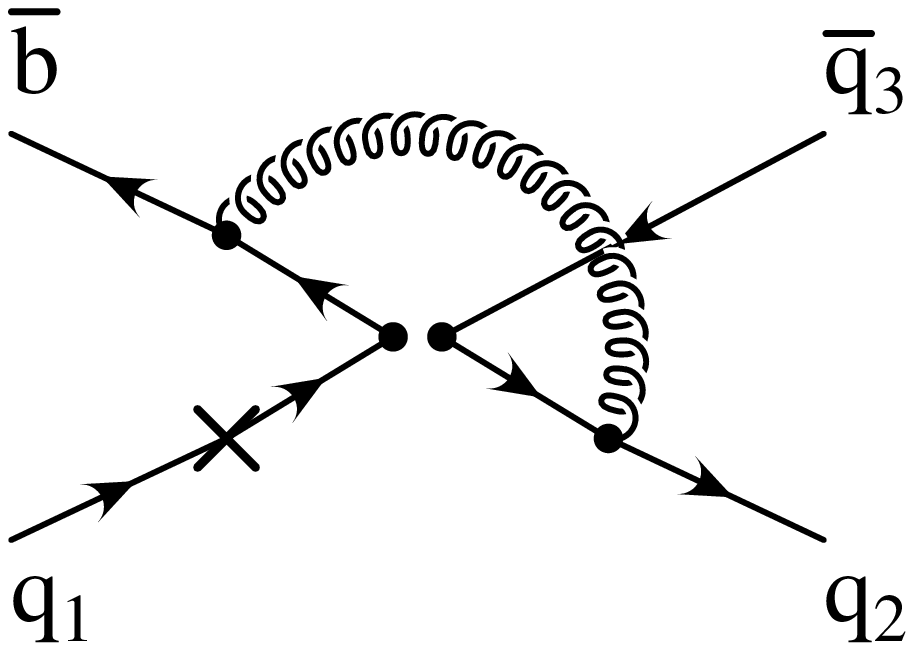,width=4cm}}\\
(a)\hspace*{5cm} (b)\\
 \mbox{\epsfig{file=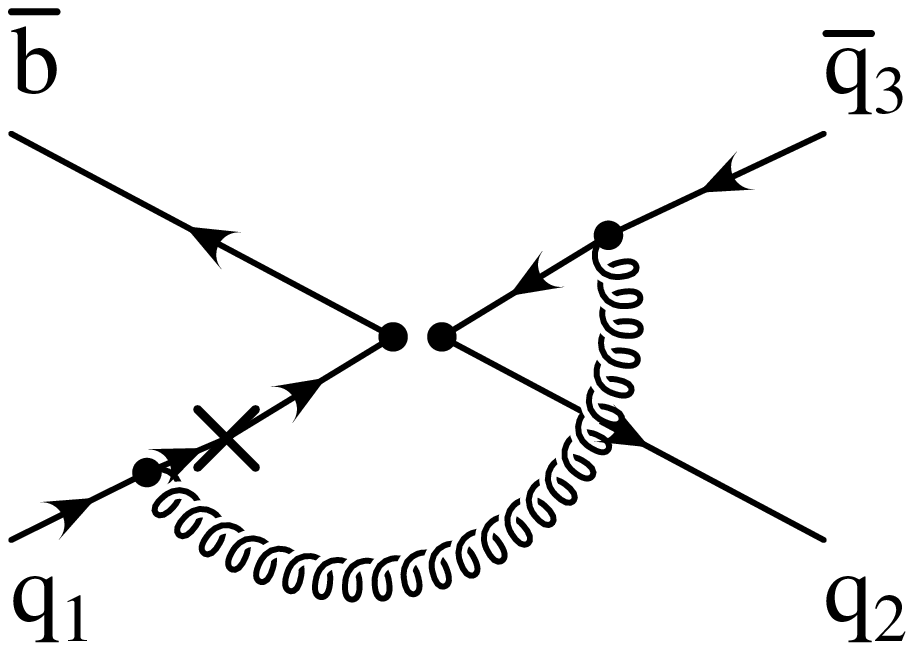,width=4cm}\qquad\qquad
 \epsfig{file=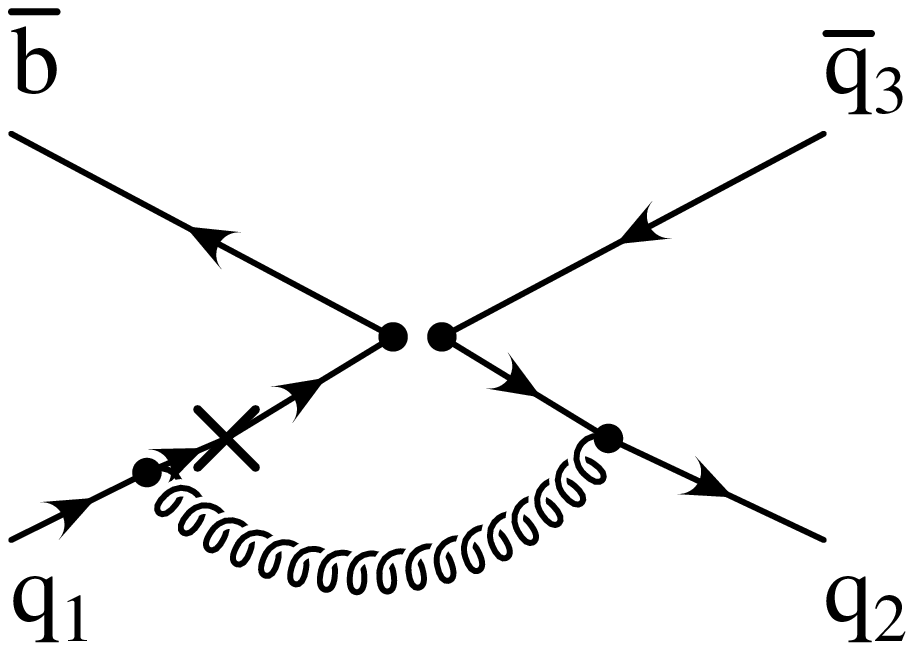,width=4cm}}\\
(c)\hspace*{5cm} (d)
 \end{center}
 \caption{
Corrections to factorization in $B\to V\gamma$ decays.
Only those diagrams are shown which contain IR divergences.
The total IR divergence cancels in their sum. The cross denotes
the attachment of the photon line. }
\label{fig2}
\end{figure}

\end{document}